\documentclass[prd,preprint,showpacs,showkeys,preprintnumbers,amsmath,amssymb,nofootinbib]{revtex4-1}

\usepackage{amsmath}
\usepackage{latexsym}

\newcommand{\be}{\begin{equation}}
\newcommand{\ee}{\end{equation}}
\newcommand{\ba}{\begin{eqnarray}}
\newcommand{\ea}{\end{eqnarray}}

\begin{document}











\title{Noncommutative Relativistic Particles}

\author{R. Amorim$^{a}$}
\email{amorim@if.ufrj.br}
\author{E. M. C. Abreu$^a$}
\email{evertonabreu@ufrrj.br}
\author{W. G. Ramirez$^{a}$}
\email{wguzmanr@if.ufrj.br}

\affiliation{$^a$Instituto de F\'{\i}sica, Universidade Federal
do Rio de Janeiro,\\
Caixa Postal 68528, 21945-970,  Rio de Janeiro, Brazil\\
${}^{b}$Grupo de F\' isica Te\'orica e Matem\'atica F\' isica, Departamento de F\'{\i}sica, Universidade Federal Rural do Rio de Janeiro\\
BR 465-07, 23890-971, Serop\'edica, RJ, Brazil\\
\today\\}


\pagestyle{myheadings}
\markright{{\it Noncommutative relativistic particles}
\hfill {\it Amorim, Abreu and Ramirez\quad p.}\hspace{1mm}}

\begin{abstract}
We present a relativistic formulation of noncommutative mechanics were the object of noncommutativity $\theta^{\mu\nu}$ is considered as an independent quantity. Its canonical conjugate momentum is also introduced, what permits to obtain an explicit form for the generators of the Lorentz group in the noncommutative case.
The  theory, which is  invariant under reparametrization, generalizes recent  nonrelativistic results. Free noncommutative bosonic particles satisfy an extended Klein-Gordon equation depending on two parameters.
\end{abstract}






\maketitle

\pagebreak

\section{Introduction }
\renewcommand{\theequation}{1.\arabic{equation}}
\setcounter{equation}{0}

\bigskip
More than sixty years ago the first paper on space-time noncommutativity was written by Snyder\cite{Snyder}. There, the space-time coordinates\footnote{ $A,B=0,1,2,3,4$; $\mu,\nu=0,1,2,3$. The parameter $a$   has dimension of length and  $\hbar=c=1$ .}  $ x^\mu$ have been promoted to operators ${\mathbf x}^{\mu}$ satisfying the  algebra

\begin{eqnarray}
\label{01}
&&[{\mathbf x}^\mu,{\mathbf x}^\nu]=
i a^2{\mathbf M}^{\mu\nu}\nonumber\\
&&[{\mathbf M}^{\mu\nu},{\mathbf x}^\lambda]=i
({\mathbf x}^{\mu}\eta^{\nu\lambda}-{\mathbf x}^{\nu}\eta^{\mu\lambda})\nonumber\\
&&[{\mathbf M}^{\mu\nu},{\mathbf M}^{\alpha\beta}]=i({\mathbf M}^{\mu\beta}\eta^{\nu\alpha}-{\mathbf M}^{\mu\alpha}\eta^{\nu\beta}+{\mathbf M}^{\nu\alpha}\eta^{\mu\beta}-
{\mathbf M}^{\nu\beta}\eta^{\mu\alpha})
\end{eqnarray}

\bigskip
\noindent which is consistent with the identification ${ \mathbf x}^{\mu}=a\,{\mathbf M}^{4\mu}$,  $M^{AB}$ representing the generators of the group $SO(1,4)$. That work was not very successfully in its original motivation, which was the introduction of a natural cutoff for quantum field theories. However, in present times, space-time noncommutativity has been a very studied subject, associated with  strings\cite{Strings} and noncommutative field theories(NCFT's)\cite{NCFT}, which are related subjects
\cite{Hull,SW}.  In NCFT's, usually  the first of relations  (\ref{01}) is  replaced by

\begin{equation}
\label{02}
[{\mathbf x}^\mu,{\mathbf x}^\nu] = i {\mathbf \theta}^{\mu\nu}
\end{equation}

\bigskip \noindent but in most situations, and contrarily to what occurs in (\ref{01}),
the object of noncommutativity  ${\mathbf \theta}^{\mu\nu}$ is considered as a constant matrix, which implies in the violation of the Lorentz symmetry\cite{NCFT}. A constant $\theta$ is indeed a consequence of the adopted  theory. When strings have their end points on D-branes, in the presence of  a constant antisymmetric tensor field background, this kind of canonical noncommutativity effectively arises. It is possible, however, to consider ${\mathbf \theta}^{\mu\nu}$  as an independent operator\cite{Carlson}, resulting in a true Lorentz invariant theory. The results of Ref.\cite{Carlson} have been applied to specific situations\cite{Haghighat,Carone,Ettefasghi} and their consequences have been explored\cite{Morita,Saxell}.
These works\cite{Carlson}-\cite{Saxell} are based on some contraction of the  algebra (\ref{01}), or equivalently, in the  so called DFR algebra\cite{DFR},
that assumes, besides (\ref{02}), the structure

\begin{eqnarray}
\label{03}
&&[{\mathbf x}^\mu,{\mathbf \theta}^{\alpha\beta}] =0\nonumber\\
&&[{\mathbf \theta}^{\mu\nu},{\mathbf \theta}^{\alpha\beta}]=0
\end{eqnarray}

\bigskip \noindent An important point of the DFR algebra is that the Weyl representation of noncommutative operators obeying (\ref{02},\ref{03})
keeps the usual form of the Moyal product, and consequently the form of the usual NCFT's, although the fields have to be considered as depending not only on ${\mathbf x}^\mu$ but also on ${\mathbf \theta}^{\alpha\beta}$.
The DFR algebra has been proposed based in arguments coming from General Relativity and Quantum Mechanics. The construction of a noncommutative theory which keeps Lorentz invariance is an important matter, since there is no experimental evidence to assume Lorentz symmetry  violation\cite{EXP}.

In noncommutative quantum mechanics\cite{Sheikh}-\cite{Rosenbaum}, as in NCFT, a similar framework with constant $\theta$ is usually employed, leading also to the violation of the Lorentz symmetry in the relativistic case or of the rotation symmetry for  nonrelativistic formulations. In two recent works
\cite{Amorim1,Amorim2} the author has explored some consequences of considering the object of noncommutativity as an independent quantity, respectively as an operator acting in Hilbert space, in the quantum case, or as a phase space coordinate, in the case of classical mechanics. In both situations it was introduced a canonical conjugate momentum for  $\theta$. It has been shown that both theories are related through  the Dirac quantization procedure, once a proper second class constraint structure is postulated. Both theories are invariant under the action of $SO(D)$.

In the present work we generalize the formalism appearing in \cite{Amorim1,Amorim2} (in its free limit) to the relativistic case, constructing in  such a way  noncommutative relativistic classical and quantum theories, both of them being invariant under the action of the Lorentz group $SO(1,D)$. As an introduction to the subject, we first present a brief review of the ordinary free relativistic particle in Section {\bf2}. In Section {\bf 3} the algebraic structure for the noncommutative case is derived, by using the Dirac theory for Hamiltonian constrained systems. The first class constraint that generates the reparametrization transformations is introduced in Section {\bf 4}. The corresponding first order action which generates the constraint structure is also  presented in that section, and its reparametrization invariance is proved. In Section {\bf 5} we present some equivalent actions, not explicitly depending on the momenta. In section {\bf 6} we discuss aspects related to the quantization of such model, where a generalized Klein-Gordon equation is derived, depending on two parameters. Concluding remarks are left for  Section {\bf 7}.

\section{The commutative relativistic particle }
\renewcommand{\theequation}{2.\arabic{equation}}
\setcounter{equation}{0}

\bigskip The commutative free relativistic particle can be described by the first order action

\begin{equation}
\label{1}
S=\int d\tau \,\,L_{FO}
\end{equation}

\bigskip \noindent where $\tau$ is an arbitrary evolution parameter and\footnote{ From this point, we adopt $\mu,\nu=0,1,2,....,D$, with arbitrary $D\geq 1$. $\eta^{\mu\nu}=diag(-1,+1,....+1)$.}

\begin{equation}
\label{2}
L_{FO}=p.\dot x-\lambda\chi
\end{equation}

\bigskip \noindent In (\ref{2}), $\dot x^\mu={{d\,x^\mu}\over{d\tau}}$, $\lambda$ is a Lagrange multiplier  and $\chi$ is a first class constraint expressing the mass shell condition

\begin{equation}
\label{3}
\chi={1\over2}(p^2+m^2)=0
\end{equation}

\bigskip The equation of motion for $p^\mu$ is just $\dot x - \lambda p=0$. If that solution is reintroduced in (\ref{1}) one obtains the einbein form of the action, where

\begin{equation}
\label{3a}
L_e={{\dot x^2}\over{2\lambda}}-{{\lambda}\over 2}m^2
\end{equation}

\bigskip\noindent
Now the equation of motion for $\lambda$ gives

\begin{equation}
\label{4}
\lambda^2=-{{\dot x^2}\over{m^2}}
\end{equation}

\bigskip \noindent and when this is introduced in $L_e$, one gets the explicit reparametrization invariant action $\int d\tau L_0$, where

\begin{equation}
\label{4a}
  L_0=-m{{\dot x^2}\over{\sqrt{-\dot x^2}}}
\end{equation}

 \bigskip \noindent All these three actions are equivalent, and are invariant under reparametrization or redefinition of the evolution parameter $\tau$.
Let us consider in some detail the first order action. Under the equal $\tau$ Poisson bracket structure   given by

\begin{equation}
\label{5}
\{x^\mu,p_\nu\}=\delta^\mu_\nu
\end{equation}

\bigskip \noindent the reparametrization invariance is generated by $G=\epsilon\chi$, where $\epsilon=\epsilon(\tau)$ is an arbitrary infinitesimal parameter and $\chi$ is given by (\ref{3}). The phase space variables $y$ are transformed accordingly to

\begin{equation}
\label{6}
\delta y= \{y,G\}
\end{equation}

\bigskip \noindent giving

\begin{eqnarray}
\label{7}
\delta x^\mu&=&\epsilon \,p^\mu\nonumber\\
\delta p_\mu&=&0
\end{eqnarray}

\bigskip As can be verified,

\begin{eqnarray}
\label{8}
\delta L_{FO}&=&p.\delta\dot x-\delta\lambda\,\chi\nonumber\\
&=&p.{{d}\over{d\tau}}\delta x-\delta\lambda\,\chi\nonumber\\
&=&\epsilon p.\dot p+ \dot \epsilon p^2 -\delta\lambda\,\chi\nonumber\\
\end{eqnarray}

\bigskip \noindent and if $\delta\lambda=\dot\epsilon$, $\delta L_{FO}$ turns in a total derivative and the variation of the action (\ref{1}) vanishes if $\epsilon$ vanishes in the extremes. This is characteristic of the so called covariant systems\cite{Dirac}.
Under quantization, the phase space variables become operators acting in Hilbert space, the brackets (\ref{5}) become commutators and this permits, for instance, that in the coordinate representation, the momenta acquire the usual derivative realization. In this situation,  the constraint (\ref{3}) acting over an state vector  gives just the Klein-Gordon equation

\begin{equation}
\label{9}
(\Box - m^2)\Psi(x)=0
\end{equation}

\bigskip \noindent which selects the physical states in Hilbert space. This guarantees that an state represented by $\Psi$ is invariant under an  unitary gauge transformation generated by $\chi$.

\section{ The noncommutative algebraic structure }
\renewcommand{\theequation}{3.\arabic{equation}}
\setcounter{equation}{0}

	\bigskip In this section,  we present a  relativist generalization of the algebraic structure  found in \cite{Amorim2}. To achieve this goal, it is introduced a  constrained Hamiltonian system  living in a phase space spanned by the quantities $x^\mu,Z^\mu$ and $\theta^{\mu\nu}$ and their conjugate momenta $p_\mu,K_\mu$ and $\pi_{\mu\nu}$. $x^\mu$ represents the usual coordinates, as in Section {\bf 2}.  $\theta^{\mu\nu}$ is the object of noncommutativity
which appears, as an operator, in (\ref{02}), and $Z^\mu$
represents auxiliary variables introduced in order to properly implement  space-time noncommutativity. After introducing the second class constraints necessary to generate the adequate Dirac brackets, $Z^\mu$ and $K_\mu$ can be
eliminated from the final results, once the constraints can be used in a strong way. Accordingly to the discussed above,
the fundamental non vanishing equal $\tau$ Poisson brackets involving all the phase space variables are given by

\begin{eqnarray}
\label{10}
\{x^\mu,p_\nu\}&=&\delta^\mu_\nu
\nonumber\\
\{\theta^{\mu\nu},\pi_{\rho\sigma}\}&=&\delta^{\mu\nu}_{\,\,\,\,\rho\sigma}\nonumber\\
\{Z^\mu,K_\nu\}&=&\delta^\mu_\nu
\end{eqnarray}

\bigskip \noindent where $\delta^{\mu\nu}_{\,\,\,\,\rho\sigma}=\delta^\mu_\rho\delta^\nu_\sigma-\delta^\mu_\sigma\delta^\nu_\rho$. The second class constraints  $\Xi^a=0, a=1,...,2D+2$, appearing in \cite{Amorim2}, are here generalized to

\begin{eqnarray}
\label{11}
\Psi^\mu&=&Z^\mu-{1\over2}\theta^{\mu\nu}p_\nu
\nonumber\\
\Phi_\mu&=&K_\mu-p_\mu
\end{eqnarray}

\bigskip \noindent  with the associated  constraint matrix


\be \label{12}
(\Delta^{ab})=  \left( \begin{array}{ll}
                               \{\Psi^\mu,\Psi^\nu\} & \{\Psi^\mu,\Phi^\nu\} \\
                               \{\Phi^\mu,\Psi^\nu\} & \{\Phi^\mu,\Phi^\nu\}
                           \end{array}
                     \right).
\ee


\bigskip \noindent with  inverse

\begin{equation}
\label{13}
(\Delta^{-1}_{ab})= \left( \begin{array}{ll}
                                \:0 & -\eta_{\mu\nu} \\
                               \eta_{\mu\nu} & \:\:\:\:0
                           \end{array}
                     \right).
\ee


 \bigskip Now the Dirac brackets between any two phase space functions $A$ and $B$ is given by \cite{Dirac}

\begin{equation}
\label{14}
\{A,B\}_D=\{A,B\}-\{A,\Xi^a\}\Delta^{-1}_{ab}\{\Xi^b,B\}
\end{equation}

\bigskip \noindent As one can verify, the algebraic structure above permits to derive the Dirac brackets

\ba
\label{15}
\{x^\mu,p_\nu\}_D & = \delta^\mu_\nu, \qquad\qquad\qquad \{x^\mu,x^\nu\}_D & = \theta^{\mu\nu} \nonumber \\
        \{p_\mu,p_\nu\}_D & = 0, \qquad\qquad\qquad \{\theta^{\mu\nu},\pi_{\rho\sigma}\}_D & = \delta^{\mu\nu}_{\,\,\,\rho\sigma}\nonumber \\
        \{\theta^{\mu\nu},\theta^{\rho\sigma}\}_D & = 0, \qquad\qquad\qquad \{\pi_{\mu\nu},\pi_{\rho\sigma}\}_D & = 0\\
        \{x^\mu,\theta^{\rho\sigma}\}_D & = 0,  \qquad\qquad\qquad \{x^\mu,\pi_{\rho\sigma}\}_D & = -{1\over2}\delta^{\mu\nu}_{\,\,\,\rho\sigma}p_\nu \nonumber \\
        \{p_\mu,\theta^{\rho\sigma}\}_D & = 0, \qquad\qquad\qquad \{p_\mu,\pi_{\rho\sigma}\}_D & = 0 \nonumber
\ea

\bigskip \noindent involving the physical variables $x^\mu, p_\mu, \theta^{\mu\nu}$ and $\pi_{\mu\nu}$. The brackets listed above   generalize the algebra found in Ref. \cite{Amorim1,Amorim2}. It is also interesting to display the remaining Dirac brackets where
the auxiliary variables $Z^\mu$ and $K_\mu$ appear:

\ba
\label{16}
\{Z^\mu,K_\nu\}_D & = 0 ,\qquad\qquad\qquad \{Z^\mu,Z^\nu\}_D & = 0 \nonumber \\
        \{K_\mu,K_\nu\}_D & = 0 ,\qquad\qquad\qquad \{Z^\mu,x^\nu\}_D & = -\,{1\over2}\,\theta^{\mu\nu} \nonumber \\
        \{K_\mu,x^\nu\}_D & = -\delta^\mu_\nu ,\qquad\qquad\qquad \{Z^\mu,p_\nu\}_D & = 0\\
        \{K_\mu,p_\nu\}_D & = 0 ,\qquad\qquad\qquad \{Z^\mu,\theta^{\sigma\rho}\}_D & = 0 \nonumber \\
        \{Z^\mu,\pi_{\sigma\rho}\}_D & = \,{1\over2}\,\delta^{\mu\nu}_{\,\,\,\sigma\rho}p_\nu ,\qquad\qquad\qquad \{K^\mu,\theta^{\sigma\rho}\}_D & = 0 \nonumber \\
        \{K_\mu,\pi_{\sigma\rho}\}_D = 0 \nonumber
\ea

\bigskip \noindent As in the nonrelativistic case,  the shifted coordinate operator

\begin{equation}
\label{17}
X^\mu=x^\mu+{1\over2}\theta^{\mu\nu}P_\nu
\end{equation}

\bigskip\noindent  also plays a fundamental role. As can be verified,

\ba
\label{18}
\{X^\mu,X^\nu\}_D=0 \qquad\qquad\,\,\,\,\,\{X^\mu,p_\nu\}_D=\delta^\mu_\nu \nonumber \\
        \{X^\mu,x^\nu\}_D={1\over2} \theta^{\mu\nu} \qquad\qquad \,\,\,\,\,\{X^\mu\theta^{\rho\sigma},\pi_{\rho\sigma}\}_D=0 \nonumber \\
        \{X^\mu,\pi_{\rho\sigma}\}_D=0 \qquad\qquad\,\,\,\,\,\{X^\mu,Z^\nu\}_D=-{1\over2} \theta^{\mu\nu} \\
        \{X^\mu,K_\nu\}_D=\delta^\mu_\nu \qquad\qquad\,\,\,\,\,\,\,\,  \nonumber
\ea

\bigskip\noindent and so the generator of the Lorentz group

\begin{equation}
\label{19}
{ J}^{\mu\nu}= { X}^\mu{ p}^\nu-{ X}^\nu{ p}^\mu-{ \theta}^{\mu\sigma}{ \pi}_\sigma^{\,\,\nu}+{ \theta}^{\nu\sigma}{ \pi}_\sigma^{\,\,\mu}
\end{equation}

\bigskip\noindent actually closes in the  $SO(1,D)$ algebra, with the use of the Dirac brackets given above. Actually

\begin{equation}
\label{20}
\{{ J}^{\mu\nu},{ J}^{\rho\sigma}\}_D=\eta^{\mu\sigma}{ J}^{\rho\nu}-\eta^{\nu\sigma}{ J}^{\rho\mu}-\eta^{\mu\rho}{ J}^{\sigma\nu}+\eta^{\nu\rho}{ J}^{\sigma\mu}
\end{equation}

\bigskip\noindent From a different point of view, a similar structure has been postulated in \cite{Gracia}.

The  Lorentz transformation of any phase space function $A$ is generated by the action of ${ J}^{\mu\nu}$. Actually, by defining

\begin{equation}
\label{21}
\delta A=-{1\over2}\epsilon_{\rho\sigma}\{A,{ J}^{\rho\sigma}\}_D
\end{equation}

\bigskip\noindent one arrives at

\begin{eqnarray}
\label{22}
\delta X^\mu&=&\epsilon ^\mu_{\,\,\nu} X^\nu\nonumber\\
\delta x^\mu&=&\epsilon ^\mu_{\,\,\nu} x^\nu\nonumber\\
\delta p_\mu&=&\epsilon _\mu^{\,\,\nu} p_\nu\nonumber\\
\delta \theta^{\mu\nu}&=&\epsilon ^\mu_{\,\,\rho} \theta^{\rho\nu}+ \epsilon ^\nu_{\,\,\rho} \theta^{\mu\rho}\nonumber\\
\delta \pi_{\mu\nu}&=&\epsilon _\mu^{\,\,\rho} \pi_{\rho\nu}+ \epsilon _\nu^{\,\,\rho} \pi_{\mu\rho}\nonumber\\
\delta Z^\mu&=&{1\over2}\epsilon ^\mu_{\,\,\nu} \theta^{\nu\rho}p_\rho\nonumber\\
\delta K_\mu&=&\epsilon _\mu^{\,\,\nu} p_\nu
\end{eqnarray}

\bigskip\noindent The last two equations are also in the proper form, once one uses the second class constraints (\ref{11}). The correct form of transformations
 (\ref{22}) guarantees the Lorentz invariance of the theory. This is only possible because of the introduction of the canonical pair
$\theta^{\mu\nu}$, $\pi_{\mu\nu}$ as independent phase space variables, which permits the existence of an object like $J^{\mu\nu}$ in (\ref{19}).

\section{The first order action }
\renewcommand{\theequation}{4.\arabic{equation}}
\setcounter{equation}{0}

This structure is almost identical to that one found in Ref.
 \cite{Amorim2}, replacing spatial indexes by
space-time indexes, and $\delta$'s by $\eta$'s in convenient places. Other points can be more subtle. Usually relativistic classical systems as relativistic particles, strings or branes, are invariant under reparametrization. This is associated with two related facts \cite{Dirac}: there are $M$ first class constraints that generate the reparametrization when the parameter space has dimension $M$, and the associated canonical Hamiltonian usually vanishes.  This is just the case treated in Section {\bf 2}, where $M=1$.
For the free noncommutative relativistic particle, this is also the case. So it is necessary to introduce
some first class constraint.
A first candidate to  be the desired  constraint is the one given by the mass shell condition (\ref{3}),
 since it has vanishing Poisson brackets with the second class constraints  (\ref{11}) and represents a suitable physical condition. One of the consequences of adopting
(\ref{3}) as the reparametrization generator is that only the physical coordinate $ x^\mu$  transforms, among all the  phase space variables.  Actually, accordingly to the prescription (\ref{6}), $G$ has vanishing Poisson brackets with all the remaining phase space variables.
The reparametrization invariance is just the invariance of the action under the redefinition of $\tau$, and there is no apparent reason to explain such an asymmetric behavior between the ordinary coordinates and the tensor ones, given by the objects of noncommutativity.

It is tempting to add to $\chi$ a term like ${1\over2}\pi^2$, but
not only its dimension is $L^{-4}$, when the dimension of $\chi$ is $L^{-2}$, as it is not first class, in the sense that it has non vanishing Poisson brackets with $\Psi^\mu$, as defined in (\ref{11}) \,\,( By construction any quantity has vanishing Dirac brackets with the second class constraints, as can be verified from (\ref{14})). A related quantity, however, is first class:

\begin{equation}
\label{23}
\chi'={1\over2}(\pi^2+K_\mu\pi^{\mu\nu}p_\nu +{1\over4}(K^2p^2-(K.p)^2))
\end{equation}

\bigskip\noindent Its form has been achieved by inspection. In the above expression  internal products are sub intended. On the second class constraint surface, however, $\chi'\approx{1\over2}\pi^2$. As can be verified,

\begin{eqnarray}
\label{24}\
\{\chi',\Psi^\mu\}&=&0\nonumber\\
\{\chi',\Phi_\mu\}&=&0
\end{eqnarray}

\bigskip\noindent   In a broad sense, the bracket structure is generated by the second class constraints and the dynamics is generated by the first class constraint, for covariant systems. If one takes the nonrelativistic limit of such a system and compare with the free limit of the Hamiltonian system  which describes the generalized noncommutative oscillator found in \cite{Amorim1,Amorim2}, it is possible to write the desired first class constraint as

\begin{equation}
\label{25}
\Upsilon={{m}\over{\Lambda}}\chi'+\chi
\end{equation}

\bigskip\noindent which, with (\ref{11}), completes the set of constraints. In (\ref{25}), $\chi$ is given by (\ref{3}) and $\chi'$ by (\ref{23}). $\Lambda$ is a parameter with dimension of $L^{-3}$, which appears in \cite{Amorim1,Amorim2}. In (\ref{25}) $m$ is not necessarily the same quantity which appears in (\ref{3}), although it has that limit for vanishing noncommutativity.

After theses points, it is possible, also in the present case, to construct an action that generates all the algebraic structure displayed above. It is written as in (\ref{1}),
but now

\begin{equation}
\label{26}
L_{FO}=p.\dot x+K.\dot Z+\pi.\dot\theta-\lambda_a\,\Xi^a-\lambda\,\Upsilon
\end{equation}

Constraints (\ref{11},\ref{25}) are generated as secondary constraints
associated with the conservation of the trivial ones that express that the canonical momenta conjugate to the Lagrange multipliers vanish identically. It is not necessary to display this procedure here since it is quite trivial. We observe that in the commutative limit where $\theta^{\mu\nu}$ and $\pi_{\mu\nu}$ vanish, $Z^\mu$ and $K_\mu$ also vanish due to (\ref{11}) and $\Upsilon$ goes to $\chi$. So (\ref{2}) is recovered from (\ref{26}).

 Now, the  reparametrization generator is assumed to be $\Upsilon$, and if one defines $G=\epsilon\,\Upsilon$,  prescription
(\ref{6}) gives

\begin{eqnarray}
\label{27}
\delta x^\mu&=&\epsilon\,[\,p^\mu+{{m}\over{2\Lambda}}(-\pi^{\mu\nu}K_\nu+
{1\over2}(K^2p^\mu-K.p\,K^\mu)\,)\,]\nonumber\\
\delta p_\mu&=&0\nonumber\\
\delta \theta^{\mu\nu}&=&\epsilon{{m}\over{\Lambda}}(\pi^{\mu\nu}-
{1\over2}(p^\mu K^\nu-p^\nu K^\mu))\nonumber\\
\delta \pi_{\mu\nu}&=&0\nonumber\\
\delta Z^\mu&=&\epsilon\,{{m}\over{2\Lambda}}[\pi^{\mu\nu}p_\nu+{1\over2}(p^2K^\mu-K.p\,p^\mu)] \nonumber\\
\delta K_\mu&=&0\nonumber\\
\end{eqnarray}

\bigskip\noindent  It is not hard to verify that under (\ref{27})

\begin{equation}
\label{28}
\delta L_{FO}=\dot\Gamma-\epsilon\dot\Upsilon-\delta\lambda\Upsilon-\delta\lambda_a\Xi^a
\end{equation}

\bigskip\noindent and so the first order action  is invariant if $\delta\lambda_a=0$ and $\delta\lambda=\dot\epsilon$,  $\epsilon$ vanishing in the extremes. In (\ref{28}),

\begin{equation}
\label{29}
\Gamma=p.\delta x+ \pi.\delta\theta+K.\delta Z
\end{equation}

\bigskip\noindent where $\delta x$, $\delta\theta$  and $\delta Z$
are given by (\ref{27}).

\section{Eliminating the momenta }
\renewcommand{\theequation}{5.\arabic{equation}}
\setcounter{equation}{0}

In the previous section the first order action has been used  to derive the  constraint structure necessary to generate the Dirac brackets  and the  reparametrization transformations. As in the ordinary case, it is also possible here to eliminate the momenta in favor of the generalized velocities and the multipliers. Let us define the modified momentum

\begin{equation}
\label{28-1}
\tilde \pi^{\alpha\beta}=\pi^{\alpha\beta}+{1\over2}(K^\alpha p^\beta-K^\beta p^\alpha)
\end{equation}

Now, the equations of motion for $p^\mu$, $K^\mu$ and $\pi^{\mu\nu}$ extracted from the first order action (\ref{26}) can be written respectively as

\begin{equation}
\label{28-2}
\dot x_\mu+\lambda_{2\mu}-\lambda p_\mu+{{m\lambda}\over{2\Lambda}}\tilde\pi_{\mu\alpha}K^\alpha-{1\over2}\theta_{\mu\alpha}\lambda_1^\alpha=0
\end{equation}

\begin{equation}
\label{28-3}
\dot Z_\mu-\lambda_{2\mu}-{{m\lambda}\over{2\Lambda}}\tilde\pi_{\mu\alpha}p^\alpha=0
\end{equation}

\bigskip\noindent and

\begin{equation}
\label{28-4}
\dot\theta_{\mu\nu}-{{m\lambda}\over{\Lambda}}\tilde\pi_{\mu\nu}=0
\end{equation}

\bigskip\noindent where $\lambda_{1\mu}$ is the Lagrange multiplier associated with the constraint $\Psi^\mu$ and $\lambda_2^\mu$ is associated with $\Phi_\mu$, as defined in (\ref{11}).
>From the above equations  formally one gets

\begin{equation}
\label{28-5}
 p_\mu={1\over\lambda}(\dot x_\mu+\dot Z_\mu -{1\over2}\theta_{\mu\nu}
\lambda_1^\mu)+\dot\theta_{\mu\alpha}\Phi^\alpha
\end{equation}

\bigskip\noindent although there is a dependence on $p$  in $\Phi$. On the second class constraint surface, however, this term vanishes since $p^\mu\approx K^\mu$. In this way
the first order Lagrangian reduces to

\begin{equation}
\label{28-6}
L_e={1\over {2\lambda}}[(\dot x+\dot Z-{1\over2}\theta.\lambda_1)^2+{\Lambda\over m}\dot\theta^2]-{\lambda\over2}m^2-\lambda_1.Z
\end{equation}

This Lagrangian also reproduce the constraint structure we are working with..
As can be verified, the first constraint in (\ref{11}) comes from the equation of motion for $\lambda_1$,\

\begin{equation}
\label{28-7}
Z^\mu-{1\over{2\lambda}}\theta^{\mu\nu}(\dot x_\nu+\dot Z_\nu-
{1\over2}\theta_{\nu\rho}\lambda_1^\rho)=0
\end{equation}

 \bigskip\noindent while the second one comes from the definition of the momenta conjugate do $x$ and to $Z$. The first class constraint comes from the equation of motion for $\lambda$, which gives a quantity that is weakly equal to (\ref{25}).
By using the equations of motion for $x^\mu$ and $Z^\mu$ as well as (\ref{28-7}), we can see that $\lambda_1^\mu$ vanishes on shell and that formally
$\dot Z^\mu={1\over{2\lambda}}(\eta-{1\over{2\lambda}}\dot\theta)^{-1}_{\mu\nu}\dot\theta^{\nu\rho}\dot x_\rho$, expression that can be introduced in (\ref{28-6}), formally eliminating the auxiliary variables, but introducing a high degree of nonlinearity.

Alternatively we can use the equation of motion of $\lambda$ to rewrite (\ref{28-6}) as

\begin{equation}
\label{28-8}
L_0=-m{{[(\dot x+\dot Z-{1\over2}\theta.\lambda_1)^2+{\Lambda\over m}\dot\theta^2]}\over{[-(\dot x+\dot Z-{1\over2}\theta.\lambda_1)^2-{\Lambda\over m}\dot\theta^2]^{1\over2}}}-\lambda_1.Z
\end{equation}

When $\theta^{\mu\nu}$ vanishes, which corresponds to the noncommutative case, (\ref{28-7}) implies that $Z^\mu$ also vanishes and both (\ref{28-6}) and (\ref{28-8}) reduce to the corresponding Lagrangians found in Section {\bf 2}. However, instead of working with these Lagrangians, we prefer to employ the first order Lagrangian (\ref{28}). As one can verify, it is simpler to implement the quantization procedure, to be worked out in the next section, by using the complete phase space since the auxiliary variables can be trivially eliminated with the use of the second class constraints.

\section{Quantization }
\renewcommand{\theequation}{6.\arabic{equation}}
\setcounter{equation}{0}

Now it is possible to quantize the classical structure displayed so far. As a first step the phase space variables $y^A$ are promoted to the operators ${\mathbf y}^A$ acting in some Hilbert space, and the Dirac quantization prescription is consistently adopted, where

\begin{equation}
\label{29.1}
\{y^A,y^B\}_D\rightarrow {1\over i}[{\mathbf y}^A,{\mathbf y}^B]
\end{equation}

As the canonical quantization is following the rule given above, all the second class constraints can be taken in a strong way. So  from (\ref{15}) it follows the equal $\tau$ commutator structure

\be \label{30}
[{\mathbf x}^\mu,{\mathbf p}_\nu]  =  i\delta^\mu_\nu, \qquad\qquad\qquad [{\mathbf x}^\mu,{\mathbf x}^\nu]  = i\,\theta^{\mu\nu} \nonumber
\ee
\be
\qquad[{\mathbf p}_\mu,{\mathbf p}_\nu]  =  0, \qquad\qquad\qquad [{\mathbf \theta}^{\mu\nu},{\mathbf \pi}_{\rho\sigma}]  = i\delta^{\mu\nu}_{\,\,\,\rho\sigma} \nonumber
\ee
\be
\!\!\!\![{\mathbf \theta}^{\mu\nu},{\mathbf \theta}^{\rho\sigma}]  = 0, \qquad\qquad\qquad [{\mathbf \pi}_{\mu\nu},{\mathbf \pi}_{\rho\sigma}]  = 0
\ee
\be
\qquad\qquad[{\mathbf x}^\mu,{\mathbf \theta}^{\rho\sigma}]  = 0, \qquad\qquad\qquad
[{\mathbf x}^\mu,{\mathbf \pi}_{\rho\sigma}]  = -{i\over2}\delta^{\mu\nu}_{\,\,\,\rho\sigma}p_\nu \nonumber
\ee
\be
[{\mathbf p}_\mu,{\mathbf \theta}^{\rho\sigma}]  = 0, \qquad\qquad\qquad [{\mathbf p}_\mu,{\mathbf \pi}_{\rho\sigma}]  = 0 \nonumber
\ee

\bigskip\noindent and it is not necessary to consider the auxiliary variables $Z^\mu$ and $K_\mu$, since the constraints (\ref{11}) are to be taken strongly. By the same reason, $\chi$' in (\ref{23}) reduces to ${1\over2}\pi^2$ and so,
the first class constraint $\Upsilon$ reduces to the simpler form

\begin{equation}
\label{31}
\Upsilon={1\over2}({\mathbf p}^2+{{m}\over{\Lambda}}{\mathbf\pi}^2+m^2)
\end{equation}

 For a theory that presents gauge degrees of freedom, the physical states are selected by imposing that they have to be annihilated by the first class constraints \cite{Dirac}. This fact assures that an unitary gauge transformation, generated by the first class constraints, keeps the physical states unchanged, as it should be. This procedure is in the foundations of several  quantization procedures of gauge theories \cite{Dirac}. In our case, if $|\Psi>$ represents a physical state  in Hilbert space, it must satisfy the condition

\begin{equation}
\label{32}
({\mathbf p}^2+{{m}\over{\Lambda}}{\mathbf\pi}^2+m^2)|\Psi>=0
\end{equation}

Observe that this constraint condition does not represent what would be obtained if we were describing a particle in a space-time with $D+1+{{D(D+1)}\over{2}}$ dimensions. This is so because $p^2$ and $\pi^2$ are independent Lorentz invariants. This anticipates the fact  that in this model the bosonic particle is classified by two parameters and not by one, given by the rest mass, as in the ordinary case.

As in the nonrelativistic case \cite{Amorim1}, it is necessary to choose a basis for the Hilbert space associated with such a system. Due to the noncommutativity between the coordinate operators, their eigenvectors can not form that basis. Again the shifted coordinate operator

\begin{equation}
\label{33}
{\mathbf X}^\mu={\mathbf x}^\mu+{1\over2}{\mathbf\theta}^{\mu\nu}{\mathbf p}_\nu
\end{equation}

\bigskip\noindent   plays a fundamental role. As one can verify,

\begin{equation}
\label{34}
\begin{array}{cc}
[{\mathbf X}^\mu,{\mathbf X}^\nu]=0 & \,\,\,\,\,[{\mathbf X}^\mu,{\mathbf p}_\nu]=i\delta^\mu_\nu\cr
   [{\mathbf X}^\mu,{\mathbf\theta}^{\rho\sigma}]=0&
     \,\,\,\,\   [{\mathbf X}^\mu,{\mathbf\pi}_{\rho\sigma}]=0\cr
\end{array}
\end{equation}

\bigskip\noindent This permits to adopt

\begin{equation}
\label{35}
{ \mathbf J}^{\mu\nu}= { \mathbf X}^\mu{\mathbf p}^\nu-{\mathbf X}^\nu{\mathbf p}^\mu-{\mathbf \theta}^{\mu\sigma}{\mathbf \pi}_\sigma^{\,\,\nu}+{\mathbf \theta}^{\nu\sigma}{\mathbf \pi}_\sigma^{\,\,\mu}
\end{equation}

\bigskip\noindent as the generators of the Lorentz group
  $SO(1,D)$, since it closes in the appropriate algebra

\begin{equation}
\label{36}
[{\mathbf J}^{\mu\nu},{\mathbf J}^{\rho\sigma}]=i\eta^{\mu\sigma}{\mathbf J}^{\rho\nu}-i\eta^{\nu\sigma}{\mathbf J}^{\rho\mu}-i\eta^{\mu\rho}{\mathbf J}^{\sigma\nu}+i\eta^{\nu\rho}{\mathbf J}^{\sigma\mu}
\end{equation}

\bigskip\noindent and generate the  Lorentz transformations, as in Section {\bf 3}, but now with the use of a commutator structure. The eigenvectors of the shifted coordinate operator (\ref{33}) also can be used in the construction of a basis in Hilbert space. Generalizing what has been done in \cite{Amorim1}, it is possible to
choose a coordinate basis $|X',\theta'>$ is such a way that

\begin{eqnarray}\label{37}
 {\mathbf X}^\mu|X',\theta'>&=&{X'}^\mu|X',\theta'>\nonumber\\
{\mathbf\theta}^{\mu\nu} |X',\theta'> &=&{\theta'}^{\mu\nu}|X',\theta'>
\end{eqnarray}

\bigskip\noindent satisfying usual orthonormality and completeness relations. In this basis

\begin{equation}
\label{38}
< { X}',{ \theta}'|{\mathbf p}_\mu|{ X}",{ \theta}">= -i{\frac{\partial}{\partial X'^\mu}}\delta^{D+1} (X'-X")\delta^{\frac{D(D+1)}{2}}({\theta}'-{\theta}")
\end{equation}

\bigskip\noindent             and

\begin{equation}
\label{39}
< { X}',{ \theta}'|{\mathbf \pi}_{\mu\nu}|{ X}",{ \theta}">= -i\delta^{D+1} (X'-X"){\frac{\partial}{\partial \theta'^{\mu\nu}}}\delta^{\frac{D(D+1)}{2}}({\theta}'-{\theta}")
\end{equation}

\bigskip\noindent implying that both momenta acquire a derivative realization.

It is interesting to redefine the variables $\theta$ and $\pi$,  by introducing the conjugate variables

\begin{eqnarray}
\label{40a}
{\mathbf Y}^{\mu\nu}&=&\sqrt{{\Lambda\over m}}{\mathbf\theta}^{\mu\nu}\nonumber\\
{\mathbf P}_{\mu\nu}&=&\sqrt{{m\over \Lambda}}{\mathbf\theta}^{\mu\nu}
\end{eqnarray}

\noindent which have respectively the same dimensions of $X^\mu$ and $P_\mu$.

Expression (\ref{34}) to (\ref{39}) are not formally modified when written in terms of the above variables. In the coordinate basis, written now in terms of the eigenvalues
${X'}^\mu$ and ${Y'}_{\mu\nu}$,  condition  (\ref{32}) is expressed as

\begin{equation}
\label{40}
(\,\Box'+ \,\,{1\over2}{{\partial}\over{\partial{Y'}^{\mu\nu}}}\,
{{\partial}\over{\partial{Y'}_{\mu\nu}}}-m^2\,)\,\Psi(X',Y')=0
\end{equation}

\bigskip\noindent where $\Box'={{\partial}\over{\partial {X'}^\mu}}\,{{\partial}\over{\partial {X'}_\mu}}$.
This extended Klein-Gordon equation is very simple. We can use the separation of variables procedure to get from (\ref{40}) the two equations

\begin{eqnarray}
\label{40b}
(\,\Box'-m^2 +\Delta)\Psi_1(X')&=&0\nonumber\\
({1\over2}\partial^{\mu\nu}\,
\partial_{\mu\nu}-\Delta))\Psi_2(Y')&=&0
\end{eqnarray}

\bigskip\noindent where $\Psi(X',Y')=\Psi_1(X')\Psi_2(Y')$,  $\partial_\mu={{\partial}\over{\partial {X'}^\mu}}$ and $\partial_{\mu\nu}={{\partial}\over{\partial {Y'}^{\mu\nu}}}$. The parameter $\Delta$ can be positive, negative or null, depending of the Lorentz character of $P_{\mu\nu}$. As commented, $|\Psi>$ depends on two parameters, $m$ and $\Delta$.
Now (\ref{40}) can be derived from the action

\begin{equation}
\label{41}
S=\int d^{D+1}\,X'\,d^{{D(D+1}\over{2}}Y'\, {1\over2}(\,\partial^\mu\Psi\partial_\mu\Psi+{{1}\over{2}} \,\,\partial^{\mu\nu}\,\Psi
\partial_{\mu\nu}\Psi-m^2\,\Psi^2)\,)
\end{equation}

\bigskip\noindent which can be taken as the starting point for implementing a second quantization procedure for the free noncommutative bosonic particle \cite{NEXT}, with interesting consequences. This modified Klein-Gordon equation and its corresponding quantum field theory can be relevant  at a high energy scale, where features characteristic of quantum gravity or string theory  probably arise. For other energy scales, the factor $\Delta$ probably can be effectively taken as a vanishing quantity, and so this approach does not imply significant modifications with respect to the  ordinary free field theory.

In a complementary point of view, it is also possible to associate the objects of noncommutativity to compactified dimensions or  to assume that the noncommutative particle is not really free but that there is some sort of confining potential in the $\theta,\pi$ sector. This  results in the introduction of a weigh function $W(\theta')$ or $W(Y')$ associated with the volume element of the corresponding action. So, in place of (\ref{41}), we could have \cite{Carlson}

\begin{equation}
\label{42}
S=\int d^{D+1}\,X'\,d^{{D(D+1}\over{2}}Y'\,W(Y') {1\over2}\,(\,\partial^\mu\Psi\partial_\mu\Psi+ \,{1\over2}\,\partial^{\mu\nu}\,\Psi
\partial_{\mu\nu}\Psi-m^2\,\Psi^2)
\end{equation}

\bigskip\noindent and the ordinary field theory action would correspond to expression (\ref{42}) after the integration in $\theta$. A similar structure arises in the nonrelativistic case \cite{Amorim1}, where there is a confining potential in the $\theta$ sector, associated with a kind of extended noncommutative oscillator, which effectively generates a weight function like $W(\theta)$.

In string theory, however, an approach similar to the one found here could present drastic consequences. This is so not only because the dynamics associated with $\theta,\pi$ could not be disconsidered, but, more important, because the counting of the bosonic degrees of freedom would be different from the one appearing in  ordinary string theory. Here the idea is that if tensor operators are included, as the objects of noncommutativity, the counting of the string bosonic degrees of freedom is not $D+1$ but $D+1+ {{D(D+1)}\over{2}}$, due to the existence of $\theta^{\mu\nu}$. This implies that in $D+1=4$, the number of bosonic degrees of freedom would be $10$. So, in a supersymmetric scheme, the string anomaly  cancelation would occur just for $D+1=4\,$. Related ideas appeared by the first time in Ref. \cite{Amorim3}, without involving noncommutativity.

\section{Conclusions}
 To close this work, we observe that it has been  possible to consistently treat the object of noncommutativity $\theta^{\mu\nu}$ as a phase space coordinate or as a Hilbert space operator, once its conjugate momentum is also considered. The classical and the corresponding quantum theory so constructed are  invariant under the action of the Lorentz group, and the results are very simple, at least in the free case. The physical states are selected by a condition that implies in   a modified Klein-Gordon equation with an extended derivative operator, involving the objects of noncommutativity. The second quantization of this model is under construction and presents interesting features \cite{NEXT}. Other point that must be considered is the introduction of interactions, for instance by using some minimal coupling procedure with extended covariant derivatives.  This program follows a route that is not the usual one found in NCFT's. Contrarily to what occurs here, the usual formulations of NCFT's do not introduce modifications in the ordinary field theories, in the free case. As it is well known \cite{NCFT}, there only interaction terms capture noncommutativity through Moyal products. These modifications  seem to be  relevant  because we expect that unusual geometrical structures may arise at very high energies, and this new physics probably should occur even for a free particle.


\vskip 1cm

\begin{thebibliography}{30}
\bibitem{Snyder} H. S. Snyder, Phys. Rev. {\bf 71} (1947)  38.
\bibitem{Strings} M. Green, J. H. Schwarz and E. Witten, {\cal  Superstring Theory},
Cambridge University Press, Cambridge,  1987; J. Polchinski, {\cal String Theory}, University Press, Cambridge, 1998; R. Szabo, {\cal An introduction to String Theory and D-Brane Dynamics}, Imperial College Press, London, 2004.
\bibitem{NCFT} R. J. Szabo, Phys. Repp {\bf378} (2003) 207.
\bibitem{Hull}M.R.Douglas and C. Hull, JHEP {\bf 9802} (1998) 008.
\bibitem{SW} N. Seiberg and E. Witten, JHEP {\bf 9909} (1999) 032.
\bibitem{Carlson} C. E. Carlson, C.D. Carone and N. Zobin, Phys. Rev. {\bf D 66} (2002) 075001.
\bibitem{Haghighat} M.  Haghighat and M. M. Ettefaghi, Phys. Rev {\bf D 70} (2004) 034017.
\bibitem{Carone} C. D. Carone and H. J. Kwee, Phys. Rev. {\bf D 73} (2006) 096005.
\bibitem{Ettefasghi} M. M. Ettefaghi and M. Haghighat, Phys. Rev {\bf D 75 } (2007) 125002.
\bibitem{Morita}H. Kase, K. Morita, Y. Okumura and E. Umezawa, Prog. Theor. Phys. {\bf 109} (2003) 663; K. Imai, K. Morita and Y. Okumura, Prog. Theor. Phys. {\bf 110} (2203) 989.
\bibitem{Saxell} S. Saxell {\it On general properties of Lorentz invariant formulation of noncommutative quantum field thery}, hep-th 08043341.
\bibitem{DFR} S. Doplicher, K. Fredenhagen and J. E. Roberts, Phys. Lett. {\bf B331} (1994) 29; Commun. Math. Phys. {\bf 172} (1995) 187.
\bibitem{EXP}J. Jaeckel, V. V. Khoze and A. Ringwald, JHEP {\bf 0602} (2006) 028.
\bibitem{Sheikh} M. M.  Sheikh-Jabbari, Phys. Lett {\bf 450} (1999) 032.
\bibitem{Durval}C. Durval and P. Horvathy, Phys. Lett. {\bf B 479} (2000) 284
\bibitem{Chaichan} M. Chaichian, M. M. Sheikh-Jabbari and A. Tureanu, Phys. Rev. Lett {\bf86} (2001) 2716.
\bibitem{Gamboa}J. Gamboa, M. Loewe and J. C. Rojas, Phys. Rev. {\bf D 64} (2001) 067901.
\bibitem{Nair}V. P .Nair and A. P. Polychronakos, Phys. Lett {\bf B 505} (2001) 267.
\bibitem{Chaichan2} M. Chaichian, A. Demichec, P. Presnajder, M. M. Sheikh-Jabbari and A. Tureanu, Nucl. Phys. {\bf B 527} (2002) 149.
\bibitem{Banerjee}R. Banerjee, Mod. Phys. Lett. {\bf 17} (2002) 631.
\bibitem{Bellucci}Stefano Bellucci and A. Nersessian, Phys. Lett. {\bf B 542} (2002) 295.
\bibitem{Ho}P.-M. Ho and H.-C. Kao, Phys. Rev Lett {\bf 88} (2002) 151602.
\bibitem{Espinosa}O. Espinosa and P. Gaete, {\cal Symmetry in noncommutative quantum mechanics}, het-th/0206066 (2002).
\bibitem{Deriglazov}A. A. Deriglazov, Phys. Lett. {\bf B555} (2003) 83; JHEP {\bf 303} (2003) 021.
\bibitem{Smailagic} A. Smailagic and E. Spallucci, J. Phys.{\bf A36} (2003) L467; J. Phys.{\bf A36} (2003) L517.
\bibitem{Jonke}L. Jonke and S. Meljanac, Eur. Phys. Jour. {\bf C29} (2003) 433.
\bibitem{Kokado}A. Kokado, T. Okamura and T. Saito, Phys. {\bf D 69} (2004) 125007.
\bibitem{Kijanka}A. Kijanka and P Kosinski, Phys. Rev. {\bf D 70} (2004) 127702.
\bibitem{Dadic}I. Dadic, L. Jonke and S. Meljanac, Acta Phys. Slov. {\bf 55} (2005) 145.
\bibitem{Bellucci1}S. Bellucci and A. Yeranyan, Phys. Lett. {\bf B 609} (2005) 418.
\bibitem{Calmet}X. Calmet, Phys. Rev. {\bf D 71} (2005) 085012; X. Calmet and M. Selvaggi, Phys. Rev {\bf D74} (2006) 037901.
\bibitem{Scholtz}F. G. Scholtz, B. Chakraborty, J. Govaerts and S. Vaidya, J. Phys. {\bf A 40} (2007) 14581.
\bibitem{Rosenbaum}M. Rosenbaum, J. David Vergara and L R. Juarez, Phys. Lett. {\bf A 267} (2007) 267.
\bibitem{Amorim1}R. Amorim, {\cal Tensor Operators in Noncommutative Quantum Mechanics}, hep-th 0804.4400.
\bibitem{Amorim2}R. Amorim, {\cal\ Tensor Coordinates in Noncommutative Mechanics}, hep-th 0804.4405.
\bibitem{Dirac}P. M. Dirac, {\cal Lectures on Quantum Mechanics}, Yeshiva University, New York, 1964; K. Sundermeyer,{\cal\  Constrained Dynamics}, Lecture Notes in Physics 169, Springer-Verlag, Berlim, 1982. M. Henneaux and C. Teitelboim, {\cal Quantization of Gauge Systems}, Princeton University Press, Princeton, 1992.
\bibitem{Gracia}J. M. Gracia-Bondía, F. Ruiz Ruiz, F. Lizzi and P. Vitale, Phys. Rev. {\bf D 74} (2006) 025014.
\bibitem{NEXT} R. Amorim, work in progress.
\bibitem{Amorim3}R. Amorim and J. Barcelos-Neto, Z. Phys. {\bf C 58} (1993) 513.
\end{thebibliography}
\end{document}